# Linear Surprisal Analysis of the H + HI → H$_2$ + I Abstraction Reaction: Further Demonstration of Kinematic Constraints on Product Energy Distributions

Benjamin Costantino
*St. Anselm's Abbey School, Washington DC, 20017*

Teresa Picconatto
*Massachusetts Institute of Technology, Cambridge, MA, 02139*

Mark Taczak
*MITRE National Security Sector, The MITRE Corporation, McLean, VA 22102*

Carl A. Picconatto
*Emerging Technologies Innovation Center, The MITRE Corporation, McLean, VA 22102*

(Dated: December 22, 2025)

**Abstract:**

Linear surprisal analysis is applied to state-to-state experimental results for the H-atom abstraction reaction, H + HI → H$_2$ + I. Contrary to previously reported results that indicated that the products from this reaction were not well fit by a linear surprisal, the reaction *can* be accurately described by linear surprisal parameters when kinematic energy constraints are taken into account. This is further evidence of the important role mass effects play in the energy disposal of state-to-state reactions and of the quantitative value of a very simple model to predict the maximum energy available to the internal states of the products.

**Introduction:**

Energy disposal in state-to-state studies has been of considerable interest for decades, leading us to an intimate understanding of the dynamics of the most basic of chemical interactions[1]. In particular, abstractions reactions, especially those involving hydrogen, have served as a testing ground for chemical dynamics and have dominated the state-to-state landscape. In many of these studies, and in particular those for which the collision energy is both substantially in excess of the barrier to reaction and a large portion of the overall reaction total energy, researchers have noted that there is a strong bias against translational energy being deposited in the internal product states.

Simple models have been offered to explain and quantify this bias[2,3], with the second being particularly successful in its ability to accurately predict the maximum internal energy of the products, as well as their internal energy distribution, for a wide range of chemical reactions[4–25], even though the exact mechanisms of the model have been shown to be overly simplistic[26]. Further, studies with reaction products that violate the energetic limits predicted by the model suggest that those results appear to come from a subset of reactions that are outside the assumptions of the model, such as those that come from orbiting, chattering, or migratory collisions[27–36].

In this work, we apply the kinematic constraints of Picconatto et al.[3] to analyze the products of the H + HI → H$_2$ + I

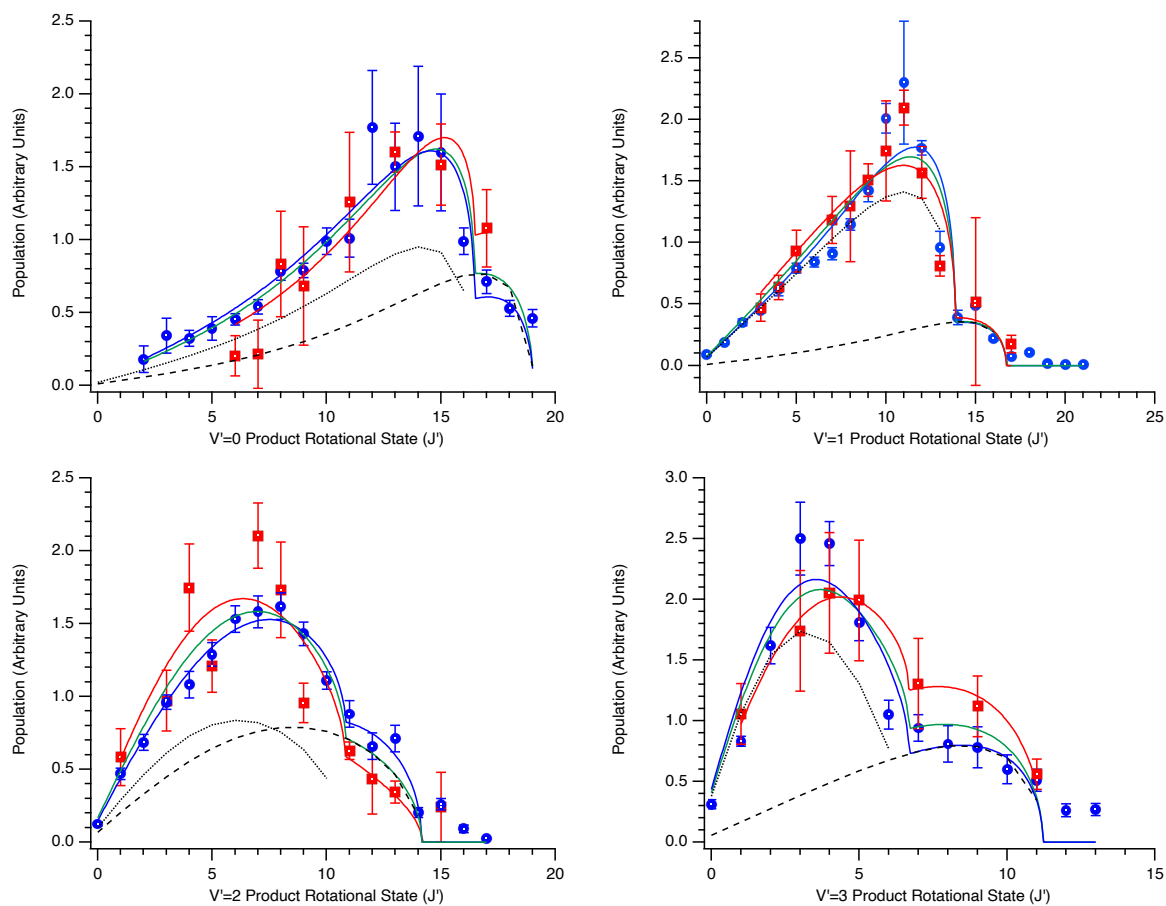

*Figure 1:* Graphs of the Rotational State Populations for the H$_2$ Products. Blue circles are from Zare; Red squares are from Valentini. Valentini data has been scaled to the Zare data for comparison purposes. Solid lines are the linear surprisal analysis of the data. Blue is Zare; Red is Valentini; green is the combined Zare and Valentini data. Dashed lines represent the deconvoluted linear surprisal analysis for the slow (small dash) and fast (large dash) H atom reactions as described in the text.

abstraction reaction. Contrary to previous assertions that the reaction products cannot be fit by a linear surprisal approach[37], we show that once the kinematic constraints are taken into account, the experimental data can be well fit by a linear surprisal, and that the results closely match calculated product distributions.[38]

**Results and Analysis:**

For light + light-heavy abstraction reactions such as the one examined here, the kinematic constraint imposes a limitation that only one-half of the collision energy is available to product internal states[3]. This is a substantial reduction in available energy, but the H + HI reaction is strongly exoergic, ($\Delta$H = -1.4 eV), so there is still a very large amount of energy available to the products, leading to the population of a significant number of rovibrational states.

As can be seen in Figure 1, we analyze the results of both of the two exhaustive single collision experimental studies of the H + HI reaction; the REMPI study of Zare et al.[39] and the CARS study from Valentini[37] and collaborators. In this figure, the Valentini data has been scaled so that the sum of the measured states has the same total value as the states that are in common with Zare, in order to make the data sets directly comparable.

In conducting this analysis, we have included both of the H atom collision energies from the dissociation of HI at 266nm that result from the formation of the I($^2P_{3/2}$) and I($^2P_{1/2}$) products. In previous experimental work[37,40], the assumption was that most, if not all, of the $H_2$ product came from collisions of the "fast" H atoms at 1.6eV total collision energy, rather than the "slow" H atoms at 0.68eV. This was empirically reasonable given the strong linear surprisal fits of all of the H + HX reactions, except for the H+HI abstraction reaction. In addition, the existing literature indicated that the "fast" H atoms were produced at twice the rate of the "slow" H atoms[41]. Combining that ratio with the relative velocities further increased the anticipated ratio of "fast" to "slow" reactions to 3:1. Finally, exchange cross sections were expected to increase with addition collision energy leading to an estimation that the 0.68eV reaction could not contribute more than 10% to the products.[37]

However, subsequent work has suggested this approximation is likely not correct. Newer work on the photodissociation of HI has measured the production ratio for the two H atoms as much closer to 1:1 and may even favor the slower atoms[42]. Further, quasi-classical trajectory calculations on the H + HI system[38] suggest that the 0.68eV reaction has a slightly larger cross section than the 1.6eV reaction. Based on the above, both channels need to be considered as they should have approximately equal weights.

This estimation is borne out by our results. As can be seen in Figure 1, the areas under the surprisal curves are approximately equal for all but the V=1 products, which appear to be dominated by the 0.68eV reaction.

In conducting our surprisal analysis, we fitted the experimental data to a double linear surprisal curve to account for both of the reaction energies.

This fit has exactly the same form as the usual surprisal approach[1], just with two terms instead of one, to account for the two reaction channels. We have fitted both of the Zare and Valentini data sets to linear surprisal curves as well as conducted a linear surprisal analysis of the combined data, by treating all experimental points as a single data set. The agreement of all of these surprisal parameters is quite good (Table 1).

| Vibrational State | | Surprisal Parameters | |
|---|---|---|---|
| | | 1.6 eV | 0.68 eV |
| V=0 | Valentini | -2.3 | -1.7 |
| | Zare | -2.4 | -1.3 |
| | Combined | -1.9 | -1.5 |
| V=1 | Valentini | -1.1 | -0.4 |
| | Zare | -1.5 | -0.9 |
| | Combined | -1.4 | -0.7 |
| V=2 | Valentini | 1.2 | 0.7 |
| | Zare | 0.0 | 0.5 |
| | Combined | 0.5 | 0.6 |
| V=3 | Valentini | 0.0 | 1.0 |
| | Zare | -0.4 | 1.2 |
| | Combined | -0.1 | 1.2 |

Table 1: Linear Surprisal Parameters for the 1.6 eV and 0.68 eV for the Valentini, Zare, and combined Valentini and Zare experimental data.

This agreement is not really surprising, as Zare has already commented on how closely his work matched that of Valentini[39]. Nevertheless, the strong overlap of the decomposed individual surprisal results for both the "fast" and "slow" H atom abstraction reactions for the two experiments is strong evidence of the importance of kinematic factors in state-to-state analysis.

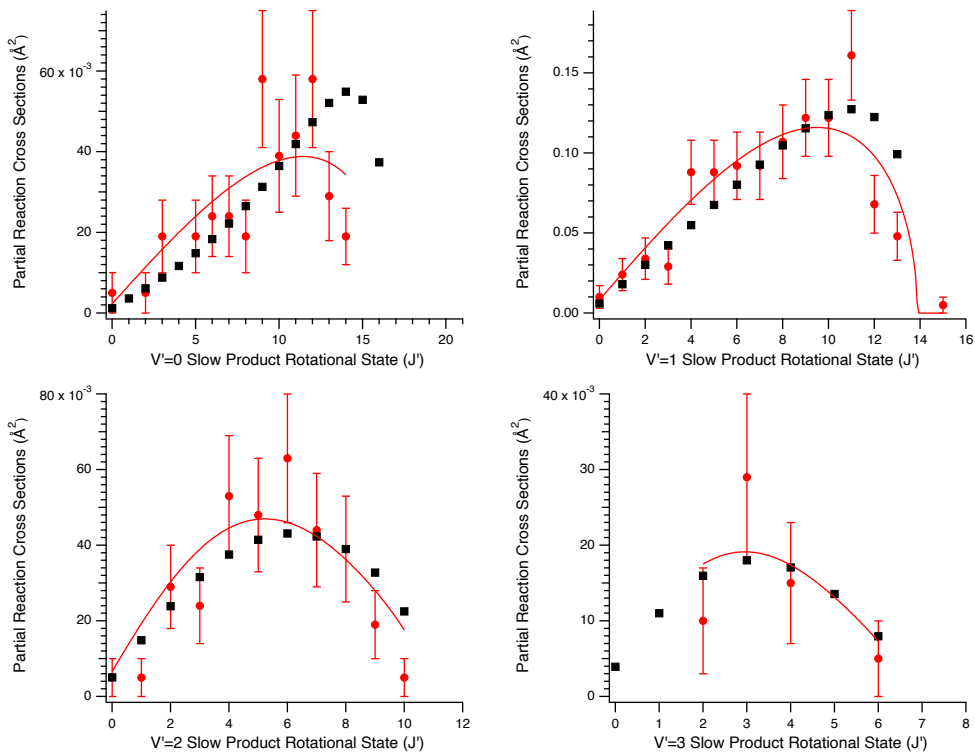

*Figure 2:* Graphs of the Rotational State Populations for the $H_2$ Products resulting from the 1.6 eV collisions as calculated by Aker and Valentini[38]. The Red circles are the QCT calculations. The Red line is a linear surprisal analysis of the QCT calculations. The Black squares are the deconvoluted linear surprisal fits from the experimental data in Figure 1.

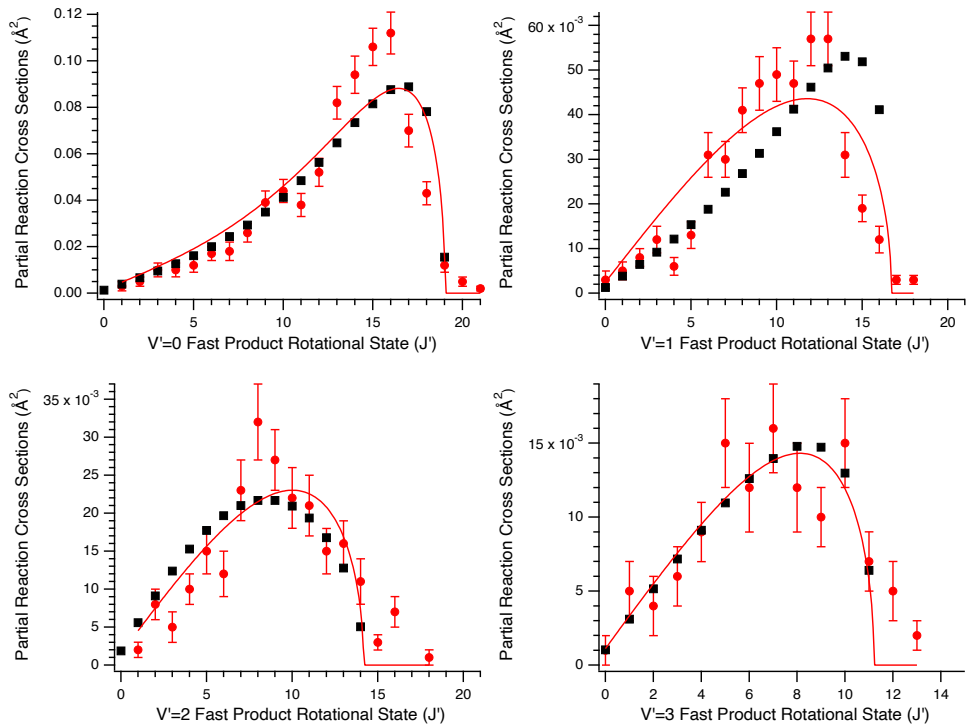

*Figure 3:* Graphs of the Rotational State Populations for the $H_2$ Products resulting from the 0.68 eV collisions as calculated by Aker and Valentini[38]. The Red circles are from QCT calculations. The Red line is a linear surprisal analysis of the QCT calculations. The Black squares are the deconvoluted linear surprisal fits from the experimental data in Figure 1.

We also compared the individual surprisal curves to the QCT results calculated by Aker and Valentini[38]. (See Figures 2 and 3.) In each of these Figures, the red circles represent the QCT calculations and the red lines represent linear surprisal fits of that data, including the kinematic constraints (performed by the present authors). The black squares are the deconvoluted individual linear surprisal fits for the "fast" and "slow" channel shown in Figure 1. As can be readily seen, the agreement is quite good. For both the "fast" and the "slow" H atom reactions, the deconvoluted linear surprisal fits from the experimental data have very strong overlap with the QCT calculation fits.

**Conclusions:**

Contrary to previous results, once kinematic constraints are taken into account, the H + HI abstraction reaction is well-fit by linear surprisal analysis. Further, the resultant experimental fits are remarkably consistent with the QCT calculations performed on react at the two, 0.68 eV and 1.6 eV, collisions energies. This analysis is further evidence of the importance of mass effects in energy disposal and utility of a simple quantitative model for predicting the total collision energy available to reaction products.